\documentclass[aps,pre,prereprint,groupeaddress,twocolumn,svgnames]{revtex4-2}
\usepackage{amsmath,amssymb,multirow,graphicx,color}
\usepackage{epsfig,amsfonts,amsbsy,color}
\usepackage{hyperref}
\usepackage{tikz}
\usetikzlibrary{shapes.arrows,fadings,shadows.blur}
\newif\ifREF
\REFtrue

\begin{document}

\title{
Violation of local equilibrium thermodynamics in one-dimensional Hamiltonian-Potts model
}

\author{Hitomi Endo}
\affiliation{
  School of Environmental Science and Engineering, Kochi University of
  Technology, Miyanoguchi 185, Tosayamada, Kami, Kochi 782-8502, Japan
}

\author{Michikazu Kobayashi}
\affiliation{
  School of Environmental Science and Engineering, Kochi University of
  Technology, Miyanoguchi 185, Tosayamada, Kami, Kochi 782-8502, Japan
}

\date{\today}

\begin{abstract}

We investigate nonequilibrium phase coexistence associated with a first-order phase transition by numerically studying a one-dimensional Hamiltonian-Potts model with fractional spatial derivatives.
The fractional derivative is introduced so as to reproduce the low-wave-number density of states of the standard
two-dimensional model, allowing phase coexistence to occur in a minimal one-dimensional setting under steady heat conduction.
By imposing a constant heat flux through boundary heat baths, we observe the stable coexistence of ordered and disordered phases separated by a stationary interface.
We find that the temperature at the interface systematically deviates from the equilibrium transition temperature,
demonstrating a clear violation of the local equilibrium description.
This deviation indicates that equilibrium metastable states can be stabilized and controlled by a steady heat current.
Furthermore, the interface temperature obtained in our simulations is in quantitative agreement with the prediction of global thermodynamics for nonequilibrium steady states.
These results confirm that the breakdown of local equilibrium and the stabilization of metastable states
are intrinsic features of nonequilibrium first-order phase transitions, independent of spatial dimensionality.
Our study thus provides a minimal and controlled numerical model for exploring the fundamental limits of thermodynamic descriptions in nonequilibrium steady states.

\end{abstract}

\maketitle

\section{Introduction}
\label{intro}
% \subsection{background}

The nonequilibrium dynamics of Hamiltonian systems are usually described by hydrodynamic equations with local equilibrium thermodynamics %
\ifREF
\cite{Landau1987Fluid,deGroot1984,Zubarev1974,McLennan1960,Sasa2014,Hayata2015}%
\fi
.
However, there are notable exceptions.
% One example is shear flow near a liquid--gas critical point%
% \ifREF
% \cite{Onuki-Kawasaki1979}%
% \fi
% , where the suppression of critical fluctuations by shear modifies the thermodynamic properties%
% \ifREF
% \cite{Onuki2004}%
% \fi
% , leading to a breakdown of local equilibrium thermodynamics.
%
One example is a first-order phase transition %
%represents a different type of singularity from that at a critical point%
\ifREF
\cite{Binder1987,Bates1988,Jin2022}%
\fi
.
A characteristic feature of first-order phase transitions is the presence of hysteresis %
\ifREF
\cite{Agarwal1981}%
\fi
. 
In an order-disorder transition %
\ifREF
\cite{Wang1989,Riste2737}%
\fi
, the transition temperature observed upon cooling from the disordered state is lower than that observed upon heating from the ordered state.
These transition temperatures, for both cooling and heating processes, generally deviate from the equilibrium transition temperature $T_{\rm c}$.
As a result, supercooled disordered states and superheated ordered states are often observed as a transient dynamical process.

These observations suggest that metastable states may become steady states when the system is driven out of equilibrium. 
To investigate this possibility, we consider steady states under heat conduction, where two heat baths with temperatures $T_1$ and $T_2$ are attached to the left and right boundaries of the system.
When the temperatures satisfy $T_1 \leq T_{\rm c} \leq T_2$, phase coexistence emerges, with ordered and disordered regions separated by an interface.

The primary goal of this paper is to determine, using a microscopic dynamical model, whether the temperature at the interface is equal to $T_{\rm c}$.
If the local equilibrium thermodynamics holds at each point in the system, the interface temperature should be equal to $T_{\rm c}$.
However, the validity of this assumption is not self-evident due to the presence of metastable
states.

Recently, a new theoretical framework, termed \textit{global thermodynamics} \cite{Nakagawa-Sasa2017,Nakagawa-Sasa2019,Nakagawa-Sasa2022}, has been proposed to describe the thermodynamics of nonequilibrium stationary states.
This theory questions the validity of the local equilibrium approximation and provides quantitative predictions for phase coexistence under a steady heat current, in contrast to other extended frameworks of thermodynamics %
\ifREF
\cite{Keizer1978,Eu1982,Jou-CasasVazquez-Lebon1988,Oono-Paniconi1998,Sasa-Tasaki2006,Bertin-Martens-Dauchot-Droz2007,Pradhan-Ramsperger-Seifert2011,Ronald2014}%
\fi
.
Global thermodynamics predicts that the interface temperature deviates from the equilibrium transition temperature, implying a breakdown of the local equilibrium approximation and suggesting that the equilibrium metastable state can be stabilized and controlled by an applied heat current.

Furthermore, this prediction is quantitatively consistent with numerical simulations of the two-dimensional Hamiltonian Potts model, which exhibits phase coexistence under a steady heat current \cite{Kobayashi-Nakagawa-Sasa2023}.
In these simulations, a heat flux is imposed at the boundaries while conserving the total energy, and phase coexistence in steady heat conduction emerges from the solution of the Hamiltonian equation of motion.
The deviation of the interface temperature from $T_{\rm c}$ is well described by the formula obtained by global thermodynamics.

However, the previous work suffers not only from finite-size effects but also from finite-roughness effects arising from finite grid spacing.
In systems with a small size and coarse spatial discretization, local equilibrium properties are restored.
This is because long-wavelength fluctuations, as well as regularities in short-wavelength fluctuations, play an essential role in the breakdown of local equilibrium.
Consequently, numerical simulations must be performed on very large spatial grids, leading to a substantial computational cost and making it difficult to carry out additional simulations and verify the reproducibility of the results.

To overcome these difficulties, we study in this paper a one-dimensional Hamiltonian Potts model with the introduction of fractional-order spatial derivatives.
In this setup, simulations can be performed with a large number of spatial grids along the direction of the heat current, since there are no transverse  directions perpendicular to it.
However, as is well known, the phase transitions do not occur in one-dimensional systems with short-range interactions.

To induce a phase transition, following a standard dimensional-reduction strategy, we introduce a fractional spatial derivative into the model so as to reproduce the same density of states at low wavenumbers as in the standard two-dimensional model.
The use of the fractional derivative provides a systematic way to reproduce the low-wave-number spectral properties of the two-dimensional model, while preserving the Hamiltonian structure, which is not straightforward in models with generic long-range interactions.
This dimensional reduction enables us to isolate the essential mechanisms responsible for the violation of local equilibrium in a minimal setting, free from geometrical constraints and finite-size effects inherent to higher-dimensional systems.
Moreover, the simplicity of the one-dimensional configuration allows for precise numerical investigations and may also facilitate experimental realizations.

By attaching heat baths with temperatures $T_1 < T_{\rm c} < T_2$ at the ends of the one-dimensional Hamiltonian Potts model, we realize steady heat conduction accompanied by phase coexistence.
We find that the interface temperature again deviates from the equilibrium transition temperature and is consistent with the prediction of global thermodynamics.
These results confirm that the stabilization of metastable states and the violation of local equilibrium are universal consequences of nonequilibrium first-order transitions, independent of system dimensionality.
Our results therefore establish a minimal model for exploring the fundamental limits of thermodynamic descriptions in nonequilibrium steady states.

The remainder of this paper is organized as follows.
In Sec. \ref{previous}, we review previous studies on the two-dimensional Hamiltonian-Potts model and global thermodynamics.
In Secs. \ref{Model} and \ref{Results}, we present the details of the one-dimensional Hamiltonian-Potts model and the results of numerical simulations, respectively.
Finally, we summarize our findings in Sec. \ref{Conclusion}.

\section{Previous work}
\label{previous}
\subsection{Overview of global thermodynamics}

The local equilibrium approximation assumes that each part of a system is locally in equilibrium and that the behavior of a nonequilibrium system can be described by integrating the properties of these local equilibrated parts %
\ifREF
\cite{Landau1987Fluid,deGroot1984,McLennan1960,Sasa2014}%
\fi
.
However, this approximation is known to break down in the presence of strong heat currents and steep temperature gradients.
In particular, when the entire system is strongly influenced by nonequilibrium effects, such as during phase transitions or in the presence of metastable states, the local equilibrium assumption fails to provide an adequate description.
% Other specific examples where the local equilibrium approximation does not hold are discussed, such as the phenomenon where critical fluctuations are suppressed when a shear flow is applied near the vapor-liquid critical point \cite{Onuki-Kawasaki1979}.

In contrast, global thermodynamics aims to characterize nonequilibrium steady states more accurately by treating the entire system as a single thermodynamic entity, rather than decomposing it into locally equilibrated subsystems \cite{Nakagawa-Sasa2017,Nakagawa-Sasa2019,Nakagawa-Sasa2022}.
% Here, a nonequilibrium steady state is a system in which macroscopic flows of heat and materials exist but there is no macroscopic time variation, and differs from an equilibrium state in that there is a flow.
A central concept in this framework is the introduction of a global temperature.
The global temperature reflects the energy distribution of the entire system and is constructed to satisfy fundamental thermodynamic relations even under nonequilibrium conditions.
Specifically, the global temperature $\tilde{T}$ is defined as the spatial average of the local temperature weighted by the density,
\begin{align}
    \tilde{T} = \frac{\displaystyle \int_{0}^{L_x} dx\: \rho(x) T(x)}{\displaystyle \int_{0}^{L_x}dx\:\rho(x)},
    \label{eq:Global-temperature}
\end{align}
where $\rho(x)$ is the density and $T(x)$ is the local temperature at position $x$, and $L_x$ is the system size.
In equilibrium statistical mechanics, temperature is defined through the average kinetic energy.
Equation~\eqref{eq:Global-temperature} can be regarded as its extension to nonequilibrium steady states, expressed as a density-weighted spatial average of the local kinetic temperature.

Furthermore, within the framework of global thermodynamics, a variational principle is formulated to determine the thermodynamic properties of phase coexistence under steady heat flow.
This variational principle can be viewed as a natural extension of Maxwell's construction in equilibrium and provides theoretical predictions for nonequilibrium steady states.
One of the key results derived from this framework is a quantitative prediction for the deviation of the interface temperature of a first-order order-disorder transition from the equilibrium transition temperature.
This result implies that a superheated ordered state can be stably maintained near the interface, offering perspectives on the thermodynamics of nonequilibrium steady states.

\subsection{Prediction by global thermodynamics}

As discussed above, global thermodynamics predicts that the interface temperature deviates from the equilibrium transition temperature.
For an order-disorder phase transition with a constant density $\rho(x) = \rho$, in Eq. \eqref{eq:Global-temperature}, the relationship between the interface temperature and the equilibrium transition temperature is given by global thermodynamics as follows \cite{Nakagawa-Sasa2017,Nakagawa-Sasa2019,Nakagawa-Sasa2022,Kobayashi-Nakagawa-Sasa2023}:
\begin{align}
    \tilde{T}_{\rm o} = \frac{T_1 + \theta}{2}, \quad \tilde{T}_{\rm d} = \frac{T_2 + \theta}{2},
    \label{prediction1}
\end{align}
which holds in the linear-response regime close to equilibrium.
Here, $X$ denotes the interface position, $\theta = T(X)$ is the interface temperature, and $T_1$ and $T_2$ are the temperatures of the heat baths satisfying $T_1 < T_{\rm c} < T_2$.
The quantities $\tilde{T}_{\rm o}$ and $\tilde{T}_{\rm d}$ represent the global temperatures of the ordered and disordered regions, respectively, and satisfy
\begin{align}
    \tilde{T} = \tilde{T}_{\rm o} \frac{X}{L_x} + \tilde{T}_{\rm d} \frac{L_x - X}{L_x},
\end{align}
where $L_x$ denotes the system size in the direction of heat conduction.
Furthermore, $\tilde{T}_{\rm o}$ and $\tilde{T}_{\rm d}$ obey
\begin{align}
    \tilde{T}_{\rm o} = \tilde{T} - \frac{\phi T_{\rm c}}{2}\frac{L_x - X}{L_x}, \quad
    \tilde{T}_{\rm d} = \tilde{T} + \frac{\phi T_{\rm c}}{2}\frac{X}{L_x},
\label{prediction2}
\end{align}
where  $\phi = (T_2 - T_1) / T_{\rm c}$ characterizes the temperature difference between the hot and cold heat baths.
By combining Eqs.~\eqref{prediction1} and \eqref{prediction2}, the interface temperature is predicted as
\begin{align}
    \theta_{\rm pred} =
    \begin{cases}
        \displaystyle 2 \tilde{T} - T_1 - \phi T_{\rm c} \frac{L_x - X}{L_x} & \text{for $T_1$}, \\
        \displaystyle 2 \tilde{T} - T_2 + \phi T_{\rm c}\frac{X}{L_x} & \text{for $T_2$}.
    \end{cases}
    \label{prediction3}
\end{align}
Introducing the heat current $J$ and the thermal conductivities $\kappa_{\rm o}$ and $\kappa_{\rm d}$ of the ordered and disordered phases, respectively, the interface temperature can be rewritten as
\begin{align}
    \theta_{\rm pred} = T_{\rm c} + |J| \left(\frac{1}{\kappa_{\rm o}} - \frac{1}{\kappa_{\rm d}} \right) \frac{X (L_x - X)}{2L_x}.
    \label{prediction4}
\end{align}
This relation predicts that the interface temperature is higher (lower) than $T_{\rm c}$ when $\kappa_{\rm o} < \kappa_{\rm d}$ ($\kappa_{\rm o} > \kappa_{\rm d}$), indicating the stabilization of a superheated (supercooled) state.
% where $\theta_{\rm o}$ ($\theta_{\rm d}$) is the prediction of the interface temperature $\theta$ from the global temperature $\tilde{T}_{\rm o}$ ($\tilde{T}_{\rm d}$).
% where $\theta$ is the interface temperature, $T_1$ and $T_2$ are the temperatures of the heat baths satisfying $T_1 < T_{\rm c} < T_2$, and $\tilde{T}_{\rm o}$ and $\tilde{T}_{\rm d}$ are the global temperatures for the ordered and disordered states.
% $X$ represents the interface position, $J$ represents the heat flow, $\kappa^o$ and $\kappa^d$ represent the thermal conductivity of the low and high temperature sides, respectively, and $\phi$ represents the temperature difference between the hot and cold heat baths.
% The second term on the right side of this equation represents the deviation between the interface temperature and the transition temperature. In other words, whether the interface temperature is higher or lower than the transition temperature depends on the difference in thermal conductivity of the two phases.
% Also, $\varepsilon$ represents the nonequilibrium degree, and the third term on the right-hand side gives the correction for large deviations from the equilibrium state.
% This term can also be calculated, but it has a very complicated form and is omitted here.

\subsection{Hamiltonian Potts model}

Attempts have been made to verify the nontrivial interface temperature predicted by global thermodynamics, as discussed in the previous section, using microscopic models.
However, violations of local equilibrium have not been confirmed in numerical simulations of particle systems for gas-liquid coexistence %
\ifREF
\cite{Rosjorde-Fossmo-Bedeaux-Kjelstrup-Hafskjold2000}%
\fi
.
This is likely because extremely large numbers of particles and system sizes are required to observe thermodynamic behavior in such simulations, and the accessible system sizes are typically too small to reveal violations of local equilibrium.
Indeed, it has been estimated that at least $10^7$ particles are necessary to detect such effects \cite{Nakano-Minami2022}.

As an alternative to particle-based simulations, several studies have employed continuous-field models.
In particular, a previous study investigated the interface temperature using numerical simulations of a continuous-field model that exhibits phase coexistence under steady heat conduction \cite{Kobayashi-Nakagawa-Sasa2023}.
Since there is no established standard model for order-parameter dynamics coupled to heat conduction, a two-dimensional $n$-state Hamiltonian-Potts model was proposed.
The Hamiltonian is given by
\begin{align}
    \mathcal{H} = \frac{1}{2} \int_D d^2 \boldsymbol{r} \left\{\sum_{i=1}^{n-1} \left[(p_i)^2 + |\boldsymbol{\nabla}q_i|^2 \right] + V(q)\right\},
    \label{eq:Hamiltonian-Potts_2D}
\end{align}
where $q_i$($i = 1, \cdots, n-1$) are field variables defined on the two-dimensional domain $D = L_x \times L_y$, $p_i$ is the momenta conjugate to $q_i$, and $|\boldsymbol{\nabla}q_i|^2$ represents the gradient energy associated with spatial variations of the fields.
The potential term $V(q)$ describes the interactions among the fields and determines the structure of local energy minima.

In the previous study, the coexistence of ordered and disordered states in an isolated system was first demonstrated through numerical integration of the Hamiltonian equations of motion.
Subsequently, by fixing the total energy and imposing heat fluxes at the boundaries, phase coexistence under steady heat conduction was realized.
A key feature of this system is that the interface temperature deviates from the equilibrium transition temperature, indicating that a metastable state of the equilibrium state is stabilized by the heat current.
Moreover, the magnitude of this deviation was found to be quantitatively consistent with the prediction of global thermodynamics given by Eq.~\eqref{prediction4}.
Nevertheless, numerical simulations of the model~\eqref{eq:Hamiltonian-Potts_2D} remain computationally demanding, suffering from finite-size effects and long simulation times required to obtain sufficient ensemble statistics in steady states.

\section{Model}

\label{Model}
\subsection{Model overview}
In this section, we introduce a one-dimensional Hamiltonian Potts model, which can be regarded as a continuous extension of the discrete Potts model \cite{PottsModel}.
In the standard continuous formulation, a second-order spatial derivative appears in the Hamiltonian, as in Eq.~\eqref{eq:Hamiltonian-Potts_2D}.
As is well known, this leads to the absence of phase transitions in one-dimensional systems.
To overcome this limitation, we introduce the fractional spatial derivative.
The Hamiltonian is as follows:
\begin{align}
\begin{aligned}
& \mathcal{H}=\int dx \: h, \\
& h = \frac{1}{2} \left\{ \sum_{i=1}^{n-1} \left[(p_i)^2 + \left( \mathcal{D} q_i \right)^2\right] + \prod_{j=1}^n Q_j \right\}.
\end{aligned}
\label{1DHamiltonPotts}
\end{align}
where $q_i$ ($i = 1, \dots, n-1$) are real scalar fields defined on the one-dimensional domain $0 \leq x \leq L$ with periodic boundary conditions $q(0) = q(L_x)$, and $p_i$ are the momentum fields conjugate to $q_i$.

The potential term is constructed from
\begin{align}
    Q_j = \sum_{i=1}^{n-1} \left(q_i - \bar{q}_{ij} \right)^2.
\end{align}
For any choice of $n$, the potential energy attains its minimum at $q_i = \bar{q}_{ij}$.
The set of vectors $\bar{q}_{ij}$ is chosen such that the distances between any pair of distinct minima are equal.
To this end, we take $\bar{q}_{ij}$ to be the coordinates of the vertices of a regular simplex in $(n-1)$ dimensions.
Specifically, they satisfy
\begin{align}
\begin{aligned}
    & \sum_{j = 1}^n \bar{q}_{ij} = 0, \\
    & \sum_{i = 1}^{n - 1} (\bar{q}_{ij})^2 = 1, \\
    & \sum_{i = 1}^{n - 1} (\bar{q}_{ij} - \bar{q}_{ij^\prime})^2 = Q_0 (1 - \delta_{jj^\prime}),
\end{aligned}
\label{eq:simplex_condition}
\end{align}
where $Q_0$ is a constant, which we fix as $Q_0 = 2n / (n - 1)$ throughout this paper.
One explicit choice of $\bar{q}_{ij}$ satisfying Eq.~\eqref{eq:simplex_condition} is given by
\begin{align}
    \bar{q}_{ij} =
    \begin{cases}
        \displaystyle \frac{\sqrt{n}}{\sqrt{n-1}} \delta_{ij} - \frac{\sqrt{n}+1}{(n-1)^{3/2}} & 1 \leqq j \leqq n - 1 \\
        \displaystyle \frac{1}{\sqrt{n - 1}} & j = n
    \end{cases}.
    \label{eq:simplex_condition2}
\end{align}

\subsection{Fractional derivatives}

% \subsubsection{Details of fractional derivatives}

The spatial derivative appearing in Eq.~\eqref{1DHamiltonPotts} is a skew-Hermitian nonlocal operator acting on $q_i$, which we denote by $\mathcal{D}$.
It is defined through its Fourier representation as
\begin{align}
\begin{aligned}
    (\mathcal{D} q)(x) = \tilde{F}_x \left[\frac{ik}{\sqrt{|k|}} F_k[q(x)] \right].
\end{aligned}
\label{absolute-half_derivative}
\end{align}
This operator does not satisfy a semigroup property, and therefore, $\mathcal{D}^2 \neq \partial_x$.

Equation~\eqref{absolute-half_derivative} is defined under periodic boundary conditions, which are naturally incorporated through the Fourier representation.
Here, $L$ denotes the system size in the numerical simulations.
The Fourier transform $F_k$ and its inverse $\tilde{F}_x$ are defined as
\begin{align}
\begin{aligned}
& F_k[q(x)] = \int_0^L dx\:q(x) e^{-ikx}, \\
& \tilde{F}_x[Q(k)] = \frac{1}{L} \sum_{k} Q(k) e^{ikx},
\end{aligned}
\label{Fourier_transformation}
\end{align}
with wave numbers $k = 2 \pi n/L$ ($n \in \mathbb{Z})$.

% \subsubsection{Introduction to fractional derivatives}
Although phase transitions do not occur in one-dimensional systems with short-ranged interactions when standard spatial derivatives are used, the introduction of the nonlocal operator $\mathcal{D}$ in Eq.~\eqref{absolute-half_derivative} effectively modifies the long-wavelength properties of the system.
In particular, the density of state at low wave numbers changes from $\rho(k) \propto k^{-1/2}$ to $\rho(k) = \mathrm{const}$, which is the same scaling as in the standard two-dimensional model.
This correspondence enables the realization of phase transitions in the present one-dimensional framework.

\subsection{Methods}
In this study, based on the one-dimensional Hamiltonian-Potts model \eqref{1DHamiltonPotts}, we investigate phase transitions and phase coexistence separated by an interface.
Since a stable interface requires a first-order phase transition, we set $n = 5$ for which the Potts model exhibits a first-order phase transition in two dimensions.
The resulting $\bar{q}_{ij}$ in Eq.~\eqref{eq:simplex_condition2} are
\begin{align}
\begin{aligned}
    & \bar{q}_{i1} = \frac{1}{8} \begin{pmatrix} 3\sqrt{5}-1 & -\sqrt{5}-1 & -\sqrt{5}-1 & -\sqrt{5}-1 \end{pmatrix}, \\
    & \bar{q}_{i2} = \frac{1}{8} \begin{pmatrix} -\sqrt{5}-1 & 3\sqrt{5}-1 & -\sqrt{5}-1 & -\sqrt{5}-1 \end{pmatrix}, \\
    & \bar{q}_{i3} = \frac{1}{8} \begin{pmatrix} -\sqrt{5}-1 & -\sqrt{5}-1 & 3\sqrt{5}-1 & -\sqrt{5}-1 \end{pmatrix}, \\
    & \bar{q}_{i4} = \frac{1}{8} \begin{pmatrix} -\sqrt{5}-1 & -\sqrt{5}-1 & -\sqrt{5}-1 & 3\sqrt{5}-1 \end{pmatrix}, \\
    & \bar{q}_{i5} = \frac{1}{2} \begin{pmatrix} 1 & 1 & 1 & 1 \end{pmatrix}.
\end{aligned}
\end{align}

% The system is discretized with the lattice spacing $\Delta x = 1/4$
% $L_x$ and the one-dimensional space is divided into $N$ points. Then the distance between points is $\Delta x = L_x / N_x$. In other words, we can express $L_x = N_x\Delta{x}$. $\Delta x = 1/4$ and the calculations were performed with different data size $N_x$ when in equilibrium. In the nonequilibrium state, the calculations were performed with a fixed data size $N_x = 1024$.

\subsubsection{Equilibrium states}

We first investigate equilibrium phase transitions in the Hamiltonian-Potts model \eqref{1DHamiltonPotts}.
Equilibrium properties are analyzed using either the Langevin equation or the Hamiltonian equation.
The system is discretized with a lattice spacing $\Delta x = 1/4$.

The Langevin dynamics is given by
\begin{align}
\begin{aligned}
    & \Delta q_{i} = p_{i} \Delta t , \quad
    \Delta p_{i} = -\left(\frac{\delta \mathcal{H}}{\delta q_{i}} + \gamma p_{i}\right)\Delta t + W_{i}, \\
    & \langle W_{i} \rangle = 0, \quad
    \langle W_{i} W_{j} \rangle = 2 \gamma k_{\rm B} T (\Delta x)(\Delta t) \delta_{i,j},
\end{aligned}
\label{eq:Langevin_eq}
\end{align}
where $T$ is the temperature and $\gamma$ is the friction coefficient.
We perform simulations with system size $L_x = 256$.
The parameters are set to $\Delta t = 0.01$ and $k_{\rm B} = \gamma = 1$.
The average of physical quantity $A(q,p)$ is calculated by a temporal average over an interval $\tau$
\begin{align}
    \langle A \rangle = \frac{1}{\tau} \int_{0}^{\tau} dt \: A(q, p).
\end{align}

The Hamiltonian dynamics is described by
\begin{align}
    \frac{\partial q_{i}}{\partial t} = p_{i}, \quad 
    \frac{\partial p_{i}}{\partial t} = - \frac{\delta \mathcal{H}}{\delta q_{i}},
    \label{eq:Hamilton}
\end{align}
where the total energy $\mathcal{H}$ is conserved.
To examine equilibrium dynamics within the Hamiltonian framework, the initial conditions are prepared using equilibrium configurations obtained from the Langevin simulations.

Using the Langevin equation \eqref{eq:Langevin_eq}, we determine the average energy density $\langle h \rangle$ as a function of the temperature $T$.
In contrast, from the Hamiltonian equations~\eqref{eq:Hamilton}, the temperature $\langle T \rangle$ is defined through the thermodynamic relation between energy density $h$ and entropy.
In the standard model with local operators, this definition is equivalent to the average kinetic energy,
\begin{align}
    \langle T \rangle = \left\langle \frac{1}{L_x}\int dx \sum_{i=1}^{4} p_i^2 \right\rangle.
\end{align}
Whether this relation holds in the present model with the nonlocal operator $\mathcal{D}$ in Eq.~\eqref{absolute-half_derivative} is not obvious.
However, since the quadratic form of the kinetic term is unchanged, we expect that the equipartition property remains valid, and thus the above relation still provides a consistent definition of temperature.

\subsubsection{nonequilibrium states}

In the previous studies of the two-dimensional Hamiltonian-Potts model \eqref{eq:Hamiltonian-Potts_2D} with standard (integer-order) spatial derivatives, nonequilibrium steady states were realized by imposing constant heat input and output at the boundaries.
In that case, the heat flow can be obtained from the time derivative of the Hamiltonian density.
Specifically, for integer-order derivatives, the Hamiltonian density satisfies the local conservation law
\begin{align}
\frac{\partial h}{\partial t} + \frac{\partial J}{\partial x} = 0,
\label{integer_continuous}
\end{align}
from which the heat current $J$ is directly given by
\begin{align}
J = - \sum_{i} p_i \frac{\partial q_i}{\partial x}.
\label{integer_heat_flow}
\end{align}

In contrast, when the spatial derivative is replaced by the nonlocal operator $\mathcal{D}$, the corresponding local conservation law does not take the simple form of Eq.~\eqref{integer_continuous}, and the heat current cannot be defined straightforwardly in this manner.
Therefore, in this study, we generate nonequilibrium steady states by attaching two heat baths with different temperatures to the boundaries of the system.

\begin{figure}[htb]
    \centering
    \begin{tikzpicture}
        \draw[fill=LightBlue](0,0)--(2,0)--(2,2)--(0,2)--cycle;
        \draw[fill=LightPink](6,0)--(6,2)--(8,2)--(8,0)--cycle;
        \shade[left color = LightBlue, right color = LightBlue!60!LightPink](1.95,0.5)--(1.95,1.5)--(4,1.5)--(4,0.5)--cycle;
        \shade[left color = LightBlue!20!LightPink, right color = LightPink](4,0.5)--(4,1.5)--(6.05,1.5)--(6.05,0.5)--cycle;
        \draw(2,1.5)--(6,1.5);
        \draw(2,0.5)--(6,0.5);
        \path(0,1.15)--(2,1.15) node[midway]{Heat bath};
        \path(0,0.85)--(2,0.85) node[midway]{$T_1 < T_{\rm c}$};
        \path(6,1.15)--(8,1.15) node[midway]{Heat bath};
        \path(6,0.85)--(8,0.85) node[midway]{$T_2 > T_{\rm c}$};
        \path(0,2)--(2,2) node[above,midway]{Region (i)};
        \path(2,2)--(6,2) node[above,midway]{Region (ii)};
        \path(6,2)--(8,2) node[above,midway]{Region (iii)};
        \path(0,0)--(2,0) node[below,midway]{Langevin Eq.};
        \path(2,0)--(6,0) node[below,midway]{Hamiltonian Eq.};
        \path(6,0)--(8,0) node[below,midway]{Langevin Eq.};
        \path(2,1.15)--(4,1.15) node[midway]{Ordered};
        \path(2,0.85)--(4,0.85) node[midway]{phase};
        \path(4,1.15)--(6,1.15) node[midway]{Disordered};
        \path(4,0.85)--(6,0.85) node[midway]{phase};
        \path(2,1.5)--(6,1.5) node[above,midway]{Interface $X$};
        \draw[dashed](4,1.5)--(4,0.5);
        \path(4,0.5) node[below]{$\theta$};
    \end{tikzpicture}
    \caption{\label{fig:situation}
        Schematic illustration of the system.
        The central region (ii) evolves under Hamiltonian dynamics, while the left and right regions (i) and (iii) are coupled to heat baths at temperatures $T_1$ and $T_2$, respectively, described by Langevin equations.
        Under steady heat conduction, phase coexistence emerges with an interface at position $X$, separating the ordered and disordered phases.
        The interface temperature is denoted by $\theta$.
    }
\end{figure}
Figure \ref{fig:situation} shows a schematic illustration of the system.
The entire system is divided into three regions:
(i) $-L_x / 2 < x < 0$, (ii) $0 \leq x \leq L_x$, and (iii) $L_x < x < 3L_x/2$.
In regions (i) and (iii), we solve the Langevin equation, while in region (ii), we solve the energy-conserving Hamiltonian equation.
The temperatures of the region (i) and (iii) are fixed at $T_1$ and $T_2$, respectively.

The Langevin dynamics in regions (i) and (iii) is given by
\begin{align}
    \begin{aligned}
        \Delta q_{i} = p_{i} \Delta t, \quad
        \Delta p_{i} = -\left(\frac{\delta \mathcal{H}}{\delta q_{i}} + \gamma p_{i} \right) \Delta t + W_{1,i}, \\
        \Delta q_{i} = p_{i} \Delta t \quad, 
        \Delta p_{i} = -\left(\frac{\delta \mathcal{H}}{\delta q_{i}} + \gamma p_{i} \right) \Delta t + W_{2,i},
    \end{aligned}
    \label{T12}
\end{align}
where the noise terms satisfy 
\begin{align}
\begin{aligned}
    & \langle W_{(1,2), i}\rangle = 0, \\
    & \langle W_{(1,2), i} W_{(1,2), j} \rangle = 2\gamma k_B T_{(1,2)} (\Delta x) (\Delta t) \delta_{i ,j}.
\end{aligned}
\end{align}
In the central region (ii), the Hamiltonian dynamics \eqref{eq:Hamilton} is solved without dissipation or noise, so that the total energy is exchanged only through the boundaries.

\section{Results}
\label{Results}

In this section, we present the results of numerical simulations of the equilibrium Langevin equation \eqref{eq:Langevin_eq}, the nonequilibrium Langevin equations \eqref{T12}, and the Hamiltonian equation \eqref{eq:Hamilton}. 
The Hamiltonian and Langevin equations are integrated using a second-order symplectic scheme and a second-order stochastic Runge-Kutta scheme, respectively.
For equilibrium simulations, periodic boundary conditions implemented via the complex discrete Fourier transform are used, whereas for nonequilibrium simulations, Neumann boundary conditions implemented via the real discrete cosine transform are employed.
% $\gamma = 1$, $k_B = 1$, and $\Delta t = 0.01$ are used for the time evolution.

\subsection{Equilibrium}
\begin{figure}[htb]
 \centering
  \includegraphics[width=0.95\linewidth]{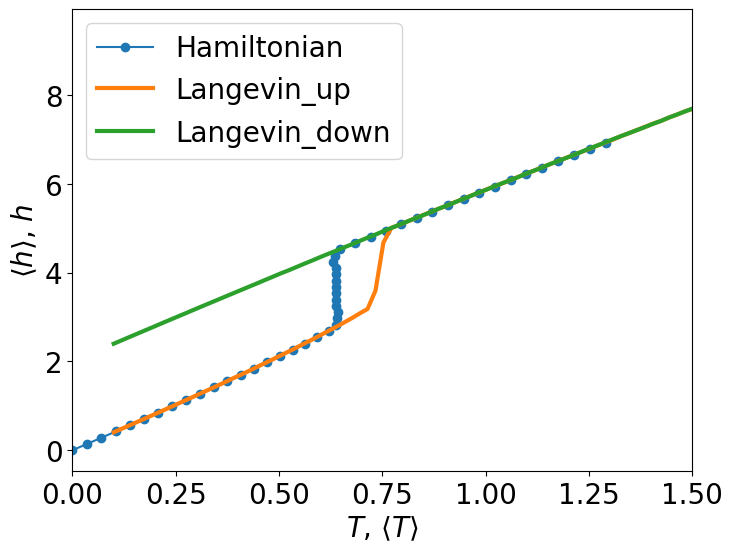}
  \caption{\label{fig:eq}
  Equilibrium properties of one-dimensional Hamiltonian-Potts model.
  The ensemble-averaged energy density $\langle h \rangle$ is shown as a function of the temperature $T$ for simulations based on the Langevin equation \eqref{eq:Langevin_eq}.
  The orange and green curves correspond to protocols in which the temperature is gradually increased from $T = 0$ and decreased from $T = 2$, respectively, revealing a clear hysteresis characteristic of a first-order phase transition.
  The blue curve shows the average temperature $\langle T \rangle$ as a function of the energy density $h$ obtained from the Hamiltonian dynamics \eqref{eq:Hamilton}, where a plateau indicates the equilibrium transition temperature.}
\end{figure}

We first present the results for equilibrium states obtained from the Langevin equation \eqref{eq:Langevin_eq} and the Hamiltonian equation \eqref{eq:Hamilton}.
Figure~\ref{fig:eq} shows the dependence of the energy density $\langle h \rangle$ on the temperature $T$.

For the Langevin dynamics, two simulation protocols are employed: the temperature is gradually increased starting from $T=0$, and gradually decreased starting from $T=2$.
As shown in Fig.~\ref{fig:eq}, the resulting hysteresis clearly indicates the presence of a first-order phase transition.

The transition temperature cannot be uniquely determined from the Langevin simulations due to the hysteresis.
In contrast, it can be identified from the Hamiltonian dynamics.
In Fig.~\ref{fig:eq}, the temperature remains nearly constant over a finite range of the energy density, yielding the equilibrium transition temperature $T_{\rm c} \simeq 0.634 \pm 0.003$.

Except in the vicinity of the transition temperature, the temperature imposed in the Langevin simulations and the temperature computed from the Hamiltonian dynamics agree closely, supporting the equivalence of the ensembles away from the phase transition.
Our results further suggest that the temperature can be consistently identified with the average kinetic energy even in the presence of the nonlocal operator.

\subsection{nonequilibrium}
We next present the results for nonequilibrium steady states obtained from the coupled dynamics of the Langevin equations~\eqref{T12} and the Hamiltonian equation~\eqref{eq:Hamilton}.
The initial conditions are prepared from equilibrium configurations of the Langevin equation~\eqref{eq:Langevin_eq} at the temperatures $T = T_1$ and $T = T_2$ in regions (i) and (iii), respectively.
Specifically, the initial fields $\{p_i^{T_1}, q_i^{T_1}\}$ and $\{p_i^{T_2}, q_i^{T_2}\}$ are sampled from the corresponding equilibrium ensemble.

In the central region (ii), the initial configurations at $t = 0$ are constructed by joining the two equilibrium configurations as
\begin{gather}
    \begin{gathered}
    p_i(0 \leq x \leq L_x/2,t=0) = p_i^{T_1}(0 \leq x \leq L_x/2),\\
    p_i(L_x/2 < x \leq L_x,t=0) = p_i^{T_2}(L_x/2 < x \leq L_x),\\
    q_i(0 \leq x \leq L_x/2,t=0) = q_i^{T_1}(0 \leq x \leq L_x/2),\\
    q_i(L_x/2 < x \leq L_x,t=0) = q_i^{T_2}(L_x/2 < x \leq L_x).
    \end{gathered}
\end{gather}

Starting from this initial state containing an interface, the system evolves in time and reaches a nonequilibrium steady state at approximately $t \simeq 1000$.
We then continue the time evolution and evaluate the time-averaged local kinetic temperature,
\begin{align}
    \langle T(x) \rangle = \left\langle \sum_{i=1}^{4} p_i^2 (x) \right\rangle.
\end{align}

\begin{figure}[htb]
  \centering
  (a) \\
  \includegraphics[width=0.95\linewidth]{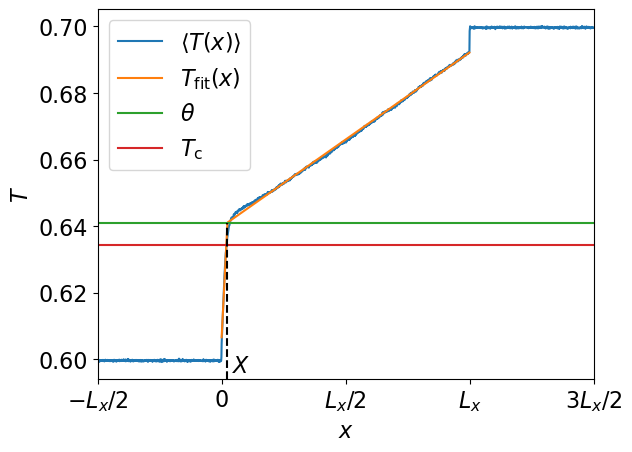} \\[5pt]
  (b) \\
  \includegraphics[width=0.95\linewidth]{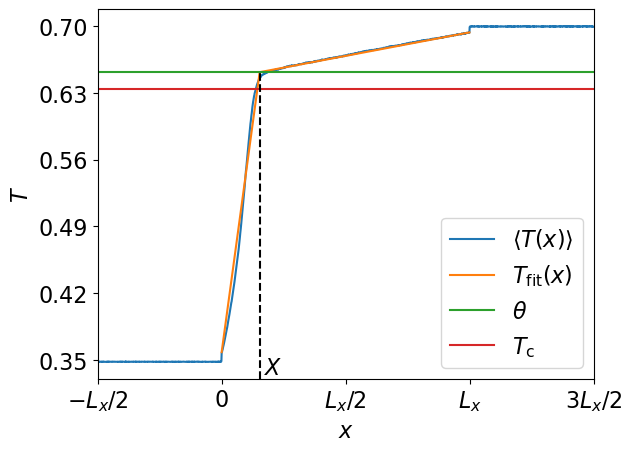}
  \caption{\label{Noneq_Tx} Spatial profiles of the local temperature $\langle T(x) \rangle$ in the nonequilibrium steady states with $T_1 = 0.6$ and $T_2 = 0.7$ [panel (a)], and $T_1 = 0.35$ and $T_2 = 0.7$ [panel (b)].
  The data in the central region $2\Delta x \leq x \leq L_x - 2\Delta x$ are fitted by the piecewise linear function $T_{\rm fit}(x)$ defined in Eq.~\eqref{T_fit}.
  The equilibrium transition temperature $T_{\rm c} \simeq 0.634$ is indicated for reference.
  The estimated interface temperatures are $\theta \simeq 0.645 \pm 6.92 \times 10^{-5}$ in panel (a) and $\theta \simeq 0.656 \pm 7.50 \times 10^{-4}$ in panel (b), both exceeding $T_{\rm c}$.}
\end{figure}

Figure~\ref{Noneq_Tx} shows the spatial profile of $\langle T(x) \rangle$ for $T_2 = 0.7$ with $T_1 = 0.6$ in panel (a) and $T_1 = 0.35$ in panel (b), both satisfying $T_1 < T_{\rm c} < T_2$.
In both cases, $\langle T(x) \rangle$ exhibits an approximately linear profile, with a clear change at the interface position $X$ separating the ordered and disordered phases.

At the boundaries between the Hamiltonian region (ii) and the Langevin heat-bath regions (i) and (iii), finite gaps in $\langle T(x) \rangle$ are observed.
To account for these boundary effects, we define the effective boundary temperatures as $T_{11} = \langle T(2\Delta x) \rangle$ and $T_{22} = \langle T(L_{x} - 2\Delta x) \rangle$.

The interface position $X$ and the interface temperature $\theta$ are determined by fitting $\langle T(x) \rangle$ to the piecewise linear function
\begin{align}
    T_{\rm fit}(x) = 
    \begin{cases}
        \displaystyle \frac{(\theta - T_{11}) (x - X)}{X - 2\Delta x} + \theta & x < X \\[10pt]
        \displaystyle \frac{(T_{22} - \theta) (x - X)}{L_x - 2\Delta x - X} + \theta & x \geq X
    \end{cases},
    \label{T_fit}
\end{align}
over the fitting range $2\Delta x \leq x \leq L_{x} - 2\Delta x$.

In both panels of Fig.~\ref{Noneq_Tx}, the interface temperature $\theta$ is clearly larger than the equilibrium transition temperature $T_{\rm c}$. This deviation indicates a breakdown of the local equilibrium description at the interface.
Furthermore, the temperature gradient in the ordered phase is steeper than that in the disordered phase, suggesting that the thermal conductivity $\kappa_{\rm o}$ in the ordered phase is smaller than $\kappa_{\rm d}$ in the disordered phase.

\begin{figure}[htb]
\centering
\includegraphics[width=0.99\linewidth]{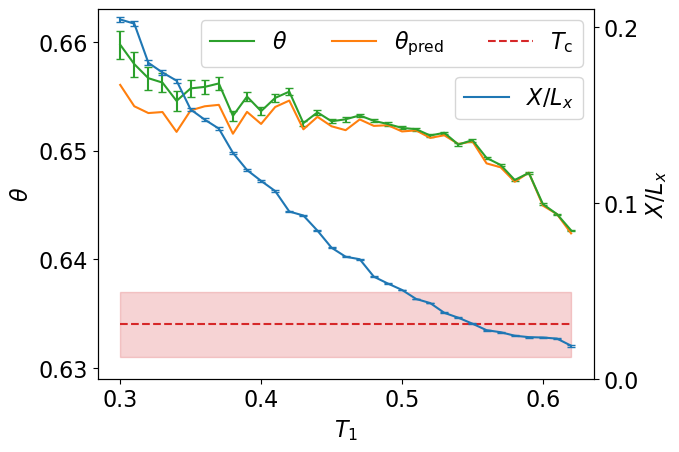}
\caption{\label{interface}
Dependence of the interface temperature $\theta$, the predicted interface temperature $\theta_{\rm pred}$ from global thermodynamics [Eq.~\eqref{prediction3}], and the interface position $X$ on the bath temperature $T_1$.
The temperature of the hot bath is fixed to $T_2 = 0.7$.
Error bars represent statistical uncertainties obtained from fitting $\langle T(x) \rangle$ with $T_{\rm fit}(x)$ in Eq.~\eqref{T_fit}.
The horizontal red line indicates the equilibrium transition temperature $T_{\rm c}$, and the shaded region represents its uncertainty.
}
\end{figure}

Figure~\ref{interface} shows the dependence of the interface temperature $\theta$ and the interface position $X$ on the bath temperature $T_1$ for fixed $T_2 = 0.7$.
The interface temperature $\theta$ exceeds $T_{\rm c}$ beyond its statistical uncertainty over the entire range of $T_1$, providing further evidence for the failure of the local equilibrium approximation at the interface.

According to global thermodynamics, the interface temperature $\theta$ is predicted as $\theta_{\rm pred}$ by Eq.~\eqref{prediction3}.
The global temperature $\tilde{T}$ is evaluated as
\begin{align}
    \tilde{T} = \frac{1}{L_x-4\Delta x} \int_{2\Delta x}^{L_x - 2\Delta x} \langle T(x) \rangle \: dx ,
\end{align}
which corresponds to the spatial average of the local kinetic temperature.
To reduce boundary effects, the integration is performed over $2\Delta x \leq x \leq L_x - 2\Delta x$ instead of the full system length.

As shown in Fig.~\ref{interface}, the agreement between $\theta$ and $\theta_{\rm pred}$ is particularly good in the range $0.4 \lesssim T_1 \lesssim T_{\rm c}$, where the system remains close to equilibrium with $T_1 \sim T_2$. This quantitative agreement supports the validity of global thermodynamics and confirms the breakdown of local equilibrium at the interface in nonequilibrium steady states.

% ここまで！！
% T11,T22を使う理由。右側で大域熱力学と一致する事実。

% The left thermal bath is set below the transition temperature, and the right thermal bath is set above it, resulting in an ordered phase on the left and a disordered phase on the right. 
% The blue line indicates the local temperature $T(x)$.
% The position and temperature of the interface were determined by a piecewise linear regression(orange lines $T_{fit}(x)$)with two lines having different slopes.
% The green line represents the estimated interface temperature $\theta$, and the red line indicates the transition temperature $T_c$.

% In all results, the interface temperature is found to be higher than the transition temperature. 
% However, since the number of samples $N_s$is relatively small in this study, the simulation results exhibit significant fluctuations and do not show smooth behavior.

% \subsection{Comparison with global thermodynamics}
% Figure \ref{比較} shows a comparison with the prediction of global thermodynamics, where the temperature of the right thermal bath is fixed at $T_2=0.7$, and the temperature of the left thermal bath is varied.

% The red line and the red shaded region represent the transition temperature$T_c$and its standard deviation, obtained from equilibrium simulations.
% The blue line indicates the interface temperature $\theta$, and the purple line shows the interface position $X_h$.
% Here,$T_o$is the global temperature on the ordered phase side, and $T_d$is the global temperature on the disordered phase side. The expression $2T_o - T_1$represents the predicted value from the low-temperature side, and $2T_d - T_2$corresponds to the predicted value from the high-temperature side.

% This graph compares the predicted interface temperature based on the global temperatures with the actual measured interface temperature. Since the predictions themselves rely on the simulation results, this cannot be considered a fully independent comparison.
% Nevertheless, the prediction that the interface temperature is higher than the transition temperature agrees with the simulation results.
% In the graph, the yellow line $2T_o - T_1$falls below the transition temperature in the low-temperature region. 
% This may be because the low-temperature side is subject to a large temperature gradient, making it a strongly nonequilibrium regime where predictions become more difficult, possibly leading to the behavior observed in Figure \ref{比較}.
% To make a proper comparison with the true prediction of global thermodynamics (Equation (\ref{予言式})), the heat current $J$ must be calculated.

\section{Conclusion}
\label{Conclusion}

In this paper, we have investigated nonequilibrium steady states associated with a first-order phase transition using a one-dimensional Hamiltonian-Potts model with fractional spatial derivatives.
By introducing a nonlocal interaction that reproduces the low-wave-number density of states of the standard two-dimensional model, we constructed a minimal setting in which phase coexistence can be realized under steady heat conduction.
This dimensional reduction enables systematic numerical simulations with large system sizes, while avoiding finite-size and geometrical effects inherent to higher-dimensional systems.

Using this model, we demonstrated that the temperature at the interface between ordered and disordered phases
deviates from the equilibrium transition temperature $T_{\rm c}$ in nonequilibrium steady states.
This deviation persists well beyond statistical uncertainty and provides clear evidence for the breakdown of the local equilibrium description at the interface.
Moreover, the observed interface temperature is in quantitative agreement with the prediction of global thermodynamics, confirming that metastable states can be stabilized by a steady heat current even in a one-dimensional system.

The present results suggest that the stabilization of metastable states and the deviation of the interface temperature from $T_{\rm c}$ are not accidental features of a specific model or dimensionality.
Rather, they are expected to be universal consequences of nonequilibrium steady states involving first-order phase transitions, independent of the spatial dimension or the detailed form of the spatial derivative.
The agreement with global thermodynamics observed here supports the view that such nonequilibrium effects are governed by macroscopic constraints rather than microscopic details.

At the same time, our study highlights the importance of carefully selecting microscopic models when investigating nonequilibrium phase coexistence.
Although the phenomenon itself is expected to be model independent, strong finite-size and discretization effects
can easily obscure the violation of local equilibrium.
The present one-dimensional fractional model provides a controlled minimal framework in which these effects are substantially reduced, allowing reliable access to the thermodynamic limit.

The predicted shift of the interface temperature should also be observable in experimental systems.
Possible candidates include gas--liquid transitions, solid--liquid transitions, and nematic transitions in liquid crystals, where steady heat currents can be imposed, and phase coexistence is well established.
Experimental verification in such systems would provide a direct test of nonequilibrium thermodynamics beyond the local equilibrium paradigm.

% From a theoretical perspective, it would be of great interest to extend the present analysis beyond the linear-response regime.
% In particular, a formulation based on the Zubarev--McLennan ensemble, which provides a systematic extension of equilibrium statistical mechanics for nonequilibrium steady states, may offer a microscopic foundation for global thermodynamics far from equilibrium.
% Such an approach could clarify the range of validity of the present predictions and deepen our understanding of thermodynamic descriptions in strongly nonequilibrium systems.

% \subsection{Future Prospects}
% \begin{itemize}
%   \item Develop new scaling relations for models based on fractional derivatives.
%   \item Perform a detailed analysis of the transition temperature in the case of first-order phase transitions.
%   \item Increase the number of samples to obtain smoother graphs.
%   \item Define the heat current $J$ and enable direct comparison with global thermodynamic predictions.
% \end{itemize}

\section*{Acknowledgment}

We would like to thank Shin-ichi Sasa and Naoko Nakagawa for the helpful suggestions and comments. This work is supported by JSPS KAKENHI Grants No. 23K22492, No. 24K00593, and No. 26K07020.

\section*{Data Availability}

The data that support the findings of this article are openly available \cite{Endo2026Data}.

% 本論文を完成させるにあたり、多くの方々のご指導とご支援をいただきました。この場をお借りして、心より感謝申し上げます。

% 本研究室の指導教官である小林未知数先生には、研究の進め方から論文の完成に至るまで、多大なご助言とご指導を賜りました。また、研究を共に進めた小林研究室の皆様には、常に励ましと協力をいただきましたこと、心より感謝申し上げます。

% 最後に、卒業までの間に関わって下さったすべての方々に、深く感謝申し上げます。

% \printbibliography
\bibliography{Endo_fig/Hamilton_Potts}
% \bibliographystyle{ChemCommun}
% \begin{thebibliography}{9}
% \bibitem{Kobayashi}
% M. Kobayashi, N. Nakagawa，S. Sasa, PRL 130, 247102 (2023)
% \bibitem{Nakagawa}
% N. Nakagawa and S. Sasa, PRL 119, 260602 (2017); JSP 177, 825 (2019); PRR 4, 033155 (2022)
% \bibitem{Ohzeki}
% M. Ohzeki and A. Ichiki, Phys. Rev. E 92, 012105 (2015).

% \end{thebibliography}

\end{document}